\documentclass[twocolumn,showpacs,preprintnumbers,amsmath,amssymb]{revtex4}

\usepackage{graphicx}
\usepackage{dcolumn}
\usepackage{bm}

\begin{document}

\preprint{APS/xxx}

\title{Scaling Invariance in Wave Functions of Quantum Systems on Complex Networks \\}
\author{Huijie Yang$^1$}
\email{huijieyangn@eyou.com}
\altaffiliation {Corresponding author}
\author{Fangcui Zhao$^2$}
\author{Yunpeng Wang$^1$}
\author{Binghong Wang$^1$}
\address{$^1$ Department of Modern Physics,
         University of Science and Technology of China,
         Anhui Hefei 230026,
         China\\
         $^2$ Life Science and Bioengineering,
         Beijing University of Technology,
         Beijing 100022,
         China }
\date{\today}

\begin{abstract}
 Structure-induced features of the wave functions for the quantum systems on
complex networks are discussed in this paper. For a quantum system
on a network, the state corresponding to the eigenvalue close to the
center of the spectrum is used as the representative state to
display the impacts of the structure on the wave functions. We
consider the Erdos-Renyi, the WS small world and the growing
randomly network (GRN) models. It is found that the probability
distribution functions (PDF) of the representative state's
components can be described with a power-law with an exponential
cutoff in a unified way. For Erdos-Renyi networks, with the increase
of the connectivity probability $p_{ER} $ the PDF turns from
power-law-dominated to exponential-dominated functions. For the WS
networks in a special region of the rewiring probability $p_r \in
(0,0.2)$, where this model can capture the features of real world
networks, and the GRN networks, the PDFs obey almost a perfect
power-law. These characteristics can be used as the structure
measurements of complex networks. They can also provide useful
information on dynamical processes on complex networks.
\end{abstract}

\maketitle

  The structures of complex networks can induce nontrivial features to the
physical processes occurring on them. Typical topics include the
epidemic spreading on networks \cite{1,2,3}, the synchronization of
coupling oscillators on networks \cite{4,5,6,7}, the response of
complex networks to stimuli \cite{8,9} and so on. In this paper, we
consider the impacts of the network structures on the wave functions
of quantum systems.

Anderson transition tells us that disorder structures can induce a
transition from extended to localized states \cite{10,11}. Quantum
systems with quasi-period structures will be in an intermediate
state, which can be described with critical wave functions
\cite{12,13}. The wave function for a localized state decreases
exponentially with the distance from its center, while the critical
wave function obeys a power-law with respect to the distance.
Complex networks have nontrivial structures rather than regular and
complete disorder ones. In literature the spectra density function
and the time series analysis methods \cite{14,15,16} are used to
reveal structure-induced features from the spectra
\cite{17,18,19,20,21,22,23} and the corresponding eigenvectors
\cite{24,25} of complex networks. In this paper we try to capture
the special characteristics of complex networks from the wave
functions of the quantum systems on them. These wave functions can
provide us useful information on the corresponding classical
systems, i.e., the probability distribution function (PDF) of
packets walking on the networks at steady states \cite{26}.
Consequently, the PDF can shed light on the dynamical processes as
the traffic flow \cite{27,28}, the epidemic spreading \cite{1,2,3}
and the synchronization on networks \cite{4,5,6,7}.

We consider an undirected complex network of $N$ coupling identical
oscillators. Denote the adjacent matrix of this network with $A$,
whose element $A_{ij}$ is $1$ and $0$ if the nodes $i$ and $j$ are
connected and disconnected, respectively. The Hamiltonian of this
quantum system reads,

\begin{equation}
\label{eq1} \hat {H} = \sum\limits_{n = 1}^N {\hat {h}_0 (x_n ,p_n
)} + \frac{1}{2}\sum\limits_{m \ne n}^N {A_{mn} \cdot \hat {V}(x_m
,x_n )} ,
\end{equation}

\noindent where $\hat {h}_0 (x_n ,p_n )$ is the Hamiltonian of the
oscillator $n$ and $\hat {V}(x_m ,x_n )$ the coupling potential
between the oscillators $m$ and $n$. Denote the site energy and the
corresponding state of each oscillator with $\varepsilon _0 $ and
$\left| {\varphi _0 } \right\rangle $, respectively. The elements of
$\hat {H}$read,

\begin{equation}
\label{eq2}
\begin{array}{l}
 H_{mn} \\
 = \left\langle {\varphi _0 (x_m )\left| {\hat {h}_0 (x_m ,p_m )\left|
{\varphi _0 (x_n )} \right.} \right.} \right\rangle \\
   + A_{mn} \cdot
   \left\langle {\varphi _0 (x_m )\left| {V(x_m ,x_n )\left| {\varphi
_0 (x_n
)} \right.} \right.} \right\rangle \\
 = \varepsilon _0 \cdot \delta _{mn} + A_{mn} \cdot V_{mn} \\
 \end{array}
\end{equation}

\noindent Assigning $\varepsilon _0 = 0$ and $V_{mn} = 1$, we have
$H = A$. Hence, the special features of the wave functions of the
considered quantum system can be obtained from the eigenvectors of
the adjacent matrix $A$.

The quantum systems considered in literature have deterministic
structures in real world Euclidean space, which lead to nature
definitions of the concepts as the localized, intermediate and
extended states of the quantum systems. These concepts are also
extended to capture the features of quantum systems on small-world
networks based on regular lattices, where the long-range edges can
be regarded in a certain degree as perturbations to the regular
lattices \cite{25}. Obviously, these concepts are invalid for
general complex networks without deterministic structures in
Euclidean space. Herein, we consider the probability distribution
functions (PDF) of the values of the components in the eigenvectors.

The eigenvector corresponding to the special eigenvalue close to the
center of the spectrum for a network, denoted with $E_c $, is
employed as the representative state to illustrate the features of
the considered quantum system. Denote the representative state with
$r = (r_1 ,r_2 , \cdots ,r_N )$, where $r_m $ is the $m$th
component. Divide the range the probability values of $\rho =
\left\{ {N \cdot \left| {r_m } \right|^2,m = 1,2,3, \cdots ,N}
\right\}$ distribute into $M$ bins. Reckon the numbers of the values
falling in the bins, respectively. The PDF of the values of the
components can be approximated as,

\begin{equation}
\label{eq3} P_t \propto K_t
\end{equation}

\noindent where $K_t $ is the number of the values falling in the
$t$th bin. The size of the bins can be chosen as a fraction of the
variance, i.e., $S_{bin} = \frac{1}{J} \cdot \sqrt
{\frac{\sum\nolimits_{m = 1}^N {\rho _m ^2} }{N}} $. In the
calculations the parameter $J$ is assigned $10$. It is found that
the PDF for different networks can be described with a power-law
with an exponential cutoff, which reads,

\begin{equation}
\label{eq4}
\begin{array}{l}
 P(T) \propto T^{ - \alpha }\exp ( - \beta T), \\
 T = t \cdot (\rho _{\max } - \rho _{\min } ) / M. \\
 \end{array}
\end{equation}

\noindent For $\alpha = 0$ and $\beta = 0$, the PDF will degenerate
to exponential and power-law functions, respectively.

The probability moment (PM) defined as,

\begin{equation}
\label{eq5} M_q (E_c ) = \sum\limits_{m = 1}^N {\left( {\frac{\rho
_m }{N}} \right)^q} ,
\end{equation}

\noindent is also used to measure the global extent of localization.
For a perfect extended state we have $M_q (E_c ) = \frac{1}{N^{q -
1}}$, while for a state strongly localized on one node the PM tends
to $1$. Generally, PM should be in the range of $\left[
{\frac{1}{N^{q - 1}},1} \right]$.

\begin{figure}
\scalebox{0.8}[0.8]{\includegraphics{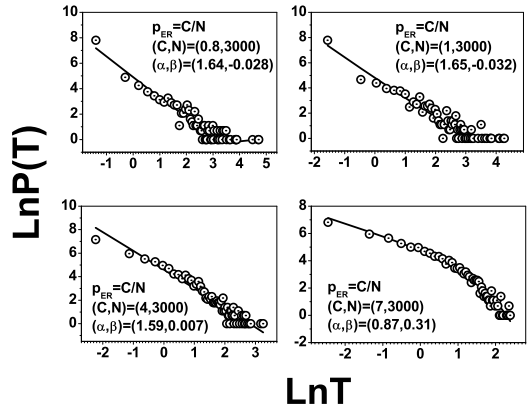}}
\caption{\label{fig:epsart} Typical $P(T)$ results for the quantum
systems on the ER networks. The circles and solid lines are the PDF
and the fitting results, respectively. The power-law with an
exponential cutoff presented in Eq.(4) can capture the
characteristics of $P(T)$ very well. The value of the parameter
$\beta $ increases from negative to positive. At the transition
point $p_{ER} = \frac{4}{N} > p_c $, we have $\beta = 0.007 \approx
0$, that is, the PDF is a perfect power-law function. When the
connectivity probability $p_{ER} = \frac{7}{N}$, $\beta $ increases
rapidly to $0.31$ while $\alpha $ decreases to $0.87$. Its PDF tends
to be exponential-dominated.}
\end{figure}

As an extreme condition, we consider networks constructed with the
Erdos-Renyi (ER) network model. An ER network can be obtained just
by connecting each pair of $N$ nodes with the connectivity
probability $p_{ER} $. There is a critical point of the connectivity
probability $p_{ER} $, denoted with $p_c $, when $p_{ER} > p_c $
almost all the nodes tend to form a giant cluster. As shown in
figure 1, with the increase of $p_{ER} $, the value of the parameter
$\beta $ turns from negative to positive. At the transition point
$p_{ER} = \frac{4}{N} > p_c $, we have $\beta = 0.007 \approx 0$.
That is, the PDF is a perfect power-law function. When the
connectivity probability $p_{ER} = \frac{7}{N}$, $\beta $ increases
rapidly to $0.31$ while $\alpha $ decreases to $0.87$. The PDF is
exponential-dominated.

\begin{figure}
\scalebox{0.8}[0.8]{\includegraphics{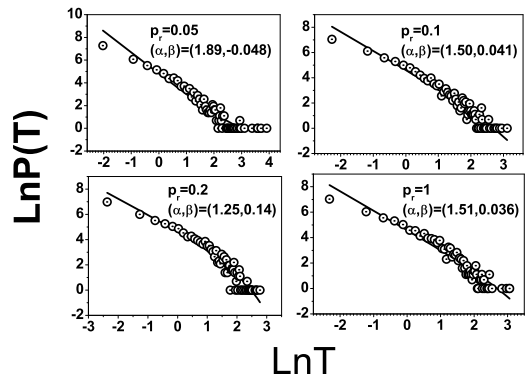}}
\caption{\label{fig:epsart}Typical$P(T)$results for the quantum
systems on the WS networks. The parameters $(N,d) = (3000,2)$. The
circles and solid lines are the PDF and the fitting results,
respectively. The power-law with an exponential cutoff presented in
Eq.(4) can capture the characteristics of $P(T)$ very well. }
\end{figure}

\begin{figure}
\scalebox{0.8}[0.8]{\includegraphics{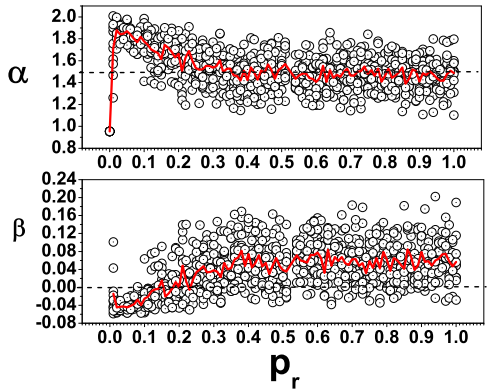}}
\caption{\label{fig:epsart}(Color online) The values of $(\alpha
,\beta )$ for the quantum systems on the constructed WS networks.
For each rewiring probability $p_r $ ten simulated results are
presented with open circles, whose averages are shown with the red
solid lines. $\left( {N,d} \right) = (3000,2)$. In the interested
region of $p_r \in (0,0.2)$ the value of $\beta $ is in the range of
$\left[ { - 0.06,0.1} \right]$. The PDFs obey almost a perfect
power-law. The corresponding values of $\alpha $ are much larger
than that in the other regions.}
\end{figure}

\begin{figure}
\scalebox{0.8}[0.8]{\includegraphics{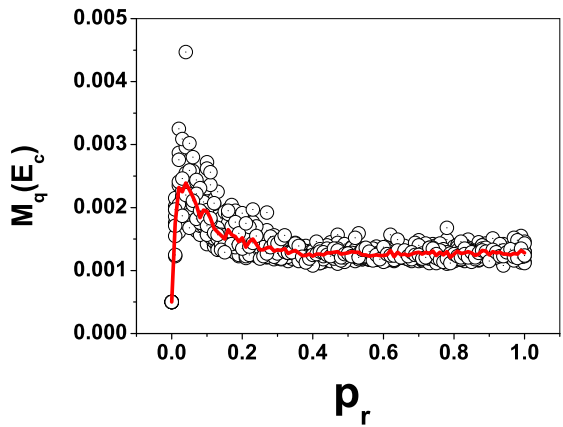}}
\caption{\label{fig:epsart}(Color online) The PM for the quantum
systems on the WS networks with different values of rewiring
probability. For each rewiring probability $p_r $ ten simulated
results are presented with open circles, whose averages are shown
with the red solid lines. $\left( {N,d} \right) = (3000,2)$. In the
region $p_r \in (0,0.2)$, the PM values are much larger than that in
the other regions. The quantum states on these networks are
localized much more than that on the other networks.}
\end{figure}

In the WS small-world model the nontrivial features are introduced
into the regular lattices just by rewiring with a certain
probability $p_r $ the end of each edge to another randomly selected
node. In the rewiring procedure double edges and self-edges are
forbidden. In the regular lattices each node is connected with $d$
right-handed nodes. Figure 2 presents several typical $P(T)$ results
for the quantum systems on the WS networks with different values of
rewiring probability. From figure 3 we can find that in the special
range of $p_r \in (0,0.2)$, where the model can capture the features
of real world networks, the values of $\beta $ are in the range of
 $\left[ { - 0.06,0.1} \right]$, i.e., the PDFs obey almost a perfect
power-law. The corresponding values of $\alpha $ are significant
larger than that in the other regions. Figure 4 presents explicitly
the higher extent of global localization of the states for the
quantum systems in the region $p_r \in (0,0.2)$. For WS networks
with $d = 5$, we can obtain similar results.

\begin{figure}
\scalebox{0.8}[0.8]{\includegraphics{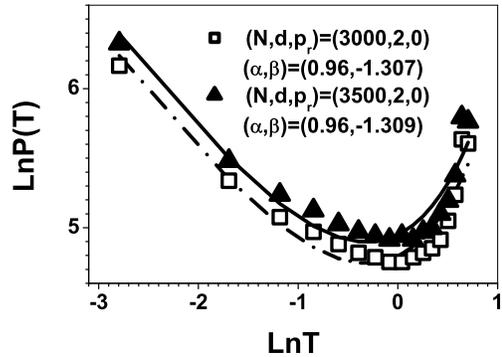}}
\caption{\label{fig:epsart}Typical $P(T)$results for quantum systems
on the regular lattices. The power-law with an exponential cutoff
presented in Eq.(4) can capture the characteristics of $P(T)$ very
well. The values of $(\alpha ,\beta )$ decrease rapidly to $(0.96, -
1.307)$. The quantum systems on these regular lattices are
significantly delocalized compared with that on WS networks (the
values of the probability $\rho $ distribute homogenously in a much
narrower interval). The circles and the solid/dashed lines are the
PDFs and the fitting results, respectively. $\left( {N,d} \right) =
(3000,2)$.}
\end{figure}

For the condition $p_r = 0$, the constructed WS networks degenerate
to one-dimensional regular lattices. The values of $(\alpha ,\beta
)$ decrease rapidly to $(0.96, - 1.307)$. The quantum systems on
these regular lattices are significantly delocalized compared with
that on WS networks (the values of the probability $\rho $
distribute homogenously in a much narrower interval).

\begin{figure}
\scalebox{0.8}[0.8]{\includegraphics{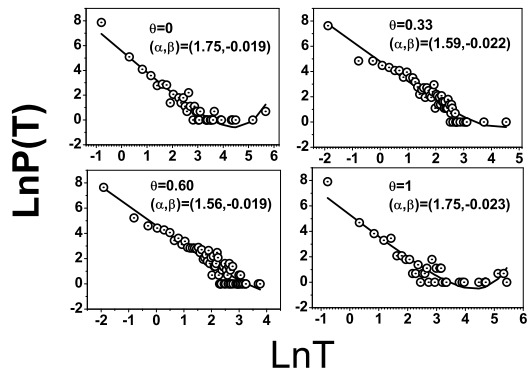}}
\caption{\label{fig:epsart}Typical$P(T)$results for the quantum
systems on the GRN networks. $(N,m) = (3000,1).$ The circles and
solid lines are the PDF and the fitting results, respectively. The
power-law with an exponential cutoff presented in Eq.(4) can capture
the characteristics of $P(T)$ very well.}
\end{figure}

The growing randomly network (GRN) model simulates a kind of
preferential attachment for new added nodes during the growing of
the networks. Starting from several connected nodes as a seed, at
time $t + 1$ a new node is linked to $m$ existing nodes at time $t$,
denoted as $\left\{ {s_t^i \left| {i = 1,2, \cdots ,m} \right.}
\right\}$, with the probability $p(s_t^i ) \propto k(s_t^i )^\theta
,0 \le \theta \le 1$. $k(s_t^i )$ is the degree of the node $s_t^i
$. $\theta = 1$ leads to the BA scale-free model. Several typical
results are shown in Figure 6. The values of $\left( {\alpha ,\beta
} \right)$ for quantum systems on the constructed GRN networks are
shown in figure 7. The values of $\beta $ are basically in the
region of $[ - 0.05,0.05]$. Consequently, the PDFs for the quantum
systems on these GRN networks obey almost a perfect power-law. The
global extents of localization are displayed in figure 8.

\begin{figure}
\scalebox{0.8}[0.8]{\includegraphics{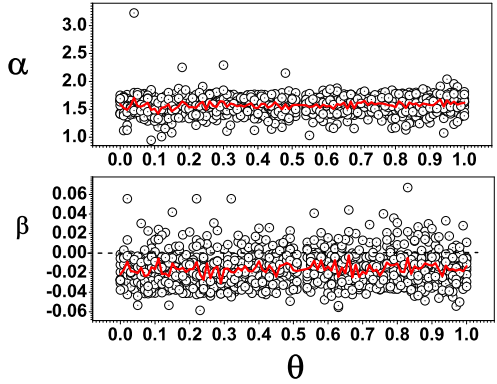}}
\caption{\label{fig:epsart}(Color online) The values of $\left(
{\alpha ,\beta } \right)$ for quantum systems on the constructed GRN
networks. The values of $\beta $ are basically in the region of $[ -
0.05,0.05]$. The PDFs for the quantum systems on these GRN networks
obey almost a perfect power-law. $\left( {N,m} \right) = (3000,1)$}
\end{figure}

\begin{figure}
\scalebox{0.8}[0.8]{\includegraphics{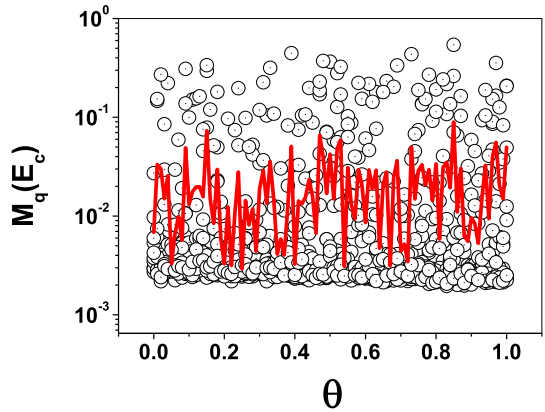}}
\caption{\label{fig:epsart}(Color online) The PM for the quantum
systems on the constructed GRN networks. Comparison with the PM
results for WS small-world networks shows that the quantum systems
on GRN networks are much more localized than that on WS small-world
networks. $\left( {N,m} \right) = (3000,1)$.}
\end{figure}

\begin{figure}
\scalebox{0.8}[0.8]{\includegraphics{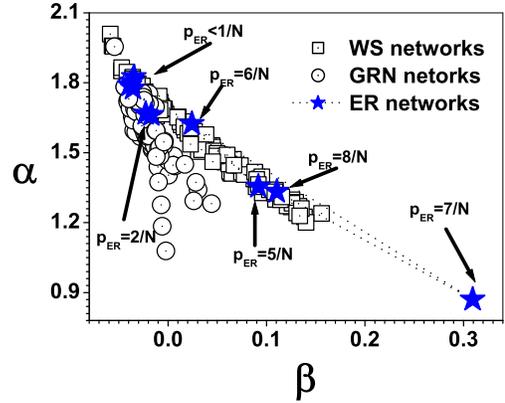}}
\caption{\label{fig:epsart}(Color online) The relation of $\alpha $
versus $\beta $. For same values of $\beta $, the $\alpha $ values
for quantum systems on the GRN networks are generally smaller than
that on the WS networks. There are much more nodes having large
values of $\rho $ in GRN networks. The ER networks with $p_{ER} $
less than or near the critical value $p_c $ ($p_{ER} \le
\frac{3}{N})$ fall into the region of GRN networks, while that with
$p_{ER} \ge \frac{4}{N}$ falls in the region of WS networks. The
network with $p_{ER} = \frac{7}{N}$ falls in the extension of the WS
network region. Its large value of $\beta = 0.31$ tells us that its
PDF is exponential-dominated.}
\end{figure}

Figure 9 presents the relation of $\alpha $ versus $\beta $. In this
scheme each point corresponds to a network. In the region of $\beta
\approx 0$ the PDF obeys perfect power-law, while a significant
deviation of $\beta $ from $0$ reveals the exponential-dominated
behavior of the PDF. For same values of $\beta $, the corresponding
$\alpha $ values for the quantum systems on the GRN networks are
generally smaller than that on the WS networks. There are much more
nodes having large values of $\rho $ in GRN networks. This higher
extent of localization can be found by comparison the PM values. The
ER networks with $p_{ER} $ less than or near the critical value $p_c
$ ($p_{ER} \le \frac{3}{N})$ fall into the region of GRN networks,
while that with $p_{ER} \ge \frac{4}{N}$ falls in the region of WS
networks. The network with $p_{ER} = \frac{7}{N}$ falls in the
extension of the WS network region. Its large value of $\beta =
0.31$ tells us that its PDF is exponential-dominated.

In summary, the states of the quantum systems on the WS, GRN and ER
networks obey a unified power-law with an exponential cutoff. The WS
networks with $p_r \in (0,0.2)$, where the WS model can capture the
features of real world networks, and the GRN networks obey almost a
perfect power-law. The values of the power-law exponent $\alpha $
for GRN networks are much smaller than that for the WS networks.
With the increase of $p_{ER} $ the PDF for ER networks tends from
almost perfect power-law to exponential-dominated functions. These
findings are consistent with the global extent of localizations for
the considered networks, which can be displayed explicitly with the
PM values.

The two parameters introduced in this paper, denoted with $\alpha $
and $\beta $, are determined by the structures of the considered
networks. Consequently, they can be employed as a measurement of the
structure characteristics. The PDF can be used to determine the
distribution of traffic packets on networks at steady states. This
information may shed light on the dynamical processes on networks,
such as the traffic flow, the traffic congestion and the epidemic
spreading processes on complex networks.

\begin{center}
\textbf{Acknowledgements}
\end{center}

This work was supported by the National Science Foundation of China
under Grant No.70571074 and No.70471033. It is also supported by the
Specialized Research Fund for the Doctoral program of Higher
Education (SRFD No. 20020358009). One of the authors (H. Yang) would
like to thank Prof. Y. Zhuo and Prof. J. Gu in China Institute of
Atomic Energy for stimulating discussions.

\end{document}